\documentclass[12pt,preprint]{aastex} 
\slugcomment{} 

\newcommand\kms{\ifmmode {\rm km\ s}^{-1} \else km s$^{-1}$\fi}
\newcommand\eflux{\ifmmode {\rm ergs\ s}^{-1}\;{\rm cm}^{-2} \else  
	ergs s$^{-1}$ cm$^{-2}$\fi}  
\newcommand\phflux{\ifmmode {\rm photons\ s}^{-1}\;{\rm cm}^{-2} 
	\else  	photons s$^{-1}$ cm$^{-2}$\fi}  
\newcommand\ergsec{\ifmmode {\rm ergs\ s}^{-1} \else  
	ergs s$^{-1}$\fi}
\newcommand\Msun{\ifmmode M_{\odot} \else $M_{\odot}$\fi} 

\def\arcsecpoint{$''\!.$}

\shorttitle{NGC3516 in the low state} 
\shortauthors{Kraemer/Turner et al. 2003}

\begin{document}
\title{Elemental Abundances in NGC 3516 } 

\author{ T.\ J.\ Turner\altaffilmark{1,2}, S.\ B.\ Kraemer\altaffilmark{3,4}, 
R.\ F.\ Mushotzky\altaffilmark{2}, I.\ M.\ George\altaffilmark{1,2}, 
J.R. Gabel\altaffilmark{3,4} }

\altaffiltext{1}{Joint Center for Astrophysics, Physics Dept., 
University of Maryland Baltimore County, 1000 Hilltop Circle, Baltimore, MD 21250}
\altaffiltext{2} {Laboratory for High Energy Astrophysics, Code 662, 
	NASA/GSFC, Greenbelt, MD 20771}
 \altaffiltext{3}{Catholic University of America, NASA/GSFC, Code 681,
Greenbelt, MD 20771}
\altaffiltext{4}{Laboratory for Astronomy and Solar Physics, Code 681, 
NASA/GSFC, Greenbelt, MD 20771}  

\begin{abstract}

We present Reflection Grating Spectrometer data from an 
{\it XMM-Newton} observation of the Seyfert 1 galaxy NGC 3516, 
taken while the continuum source was in an extremely 
low flux state.  This observation offers a rare opportunity for a 
detailed study of emission from a Seyfert 1 galaxy 
as these are usually dominated by 
 high nuclear continuum levels 
and heavy absorption. 
The spectrum shows numerous  narrow
emission lines (FWHM $\lesssim$ 1300 km s$^{-1}$) in the 0.3 -- 2 keV range,  
including the H-like lines of C, N, and O 
and the He-like lines of N, O and Ne. The emission-line ratios
and the narrow width of the radiative recombination
continuum of C{\sc vi} indicate that the gas is photoionized and
of fairly low temperature ($kT$ $\lesssim$ 0.01 keV). The availability of 
emission 
lines from different elements for two iso-electronic sequences
allows us to constrain the element abundances. These data show 
that the N lines are far stronger than would be expected from gas of
solar abundances. Based on our photoionization models we find that
nitrogen is overabundant in the central regions of the galaxy, 
compared to carbon, oxygen and neon
by at least a factor of 2.5. We suggest that this is the result of
secondary production of nitrogen in intermediate mass stars, and
indicative of the history of star formation in NGC 3516.

	\end{abstract}


	\section{Introduction}

X-ray and UV spectra of the Seyfert~1 galaxy NGC~3516 
(z=0.008836$\pm0.000023$, \citealt{keel96}) 
have shown evidence for a significant column of ionized gas 
along the line-of-sight to the nucleus 
(\citealt{kriss96,netzer02,kraemer02}). 
A recent {\it Chandra} observation of this source 
 found it to exist 
at a historical low-state. The spectrum at that epoch, and comparison with 
previous X-ray data led \citet{netzer02} to suggest a model for 
NGC~3516 where a constant column of gas reacts to changes in 
the nuclear ionizing flux. That gas was constrained to have density 
$> 2.4 \times 10^6$cm$^{-3}$ and exist at a distance 
 $< 6 \times 10^{17}h^{-2}_{75}$ cm from the nucleus. \citet{netzer02} 
also detected a strong O{\sc vii} 0.561 keV line and marginally detected 
a N{\sc vi} 0.419 keV line. Line measurements from the {\it Chandra} 
Low Energy Transmission Grating (LETG) data appeared 
consistent with  emission from the X-ray absorber. 

We present here {\it XMM-Newton} (hereafter {\it XMM}) RGS grating spectra 
of NGC~3516. These data allow us to rethink the origin of the X-ray line 
emission in NGC~3516.

	\section{The {\it XMM} RGS Observation}

An {\it XMM} 
observation  of the Seyfert 1 galaxy NGC~3516 
was performed covering November 09 UT 23:12:51 -- 11 UT 
10:54:19. This observation was part of a multi-satellite campaign which 
including overlapping observations by {\it RXTE} and {\it Chandra}. 
The Fe K$\alpha$ line shows interesting structure and evolution 
during the observation, and the combined data in the hard X-ray regime 
are detailed  by \citet{turner02}. 
Here we present a detailed analysis of the {\it XMM} RGS grating data 
from 2001 November. 

As noted by \citet{turner02} NGC~3516 
had a flux 
$F_{2-10\ keV} \sim  1.3 - 1.5 \times 10^{-11} {\rm erg\ cm^{-2}\ s^{-1}}$, 
during the November 2001 observations. 
The source was in this flux state during the previous 
{\it Chandra} LETG observation 
\citep{netzer02} and as observed by {\it ASCA} during 1999 (Figure~1). 
Unfortunately no useful RGS data were obtained from an earlier 
epoch observation with {\it XMM} in 2001 April, due to background flares 
during a period of high solar activity. 

 {\it RGS} data were processed using the SAS  5.3.3. version of 
{\sc rgsproc} and  spectra were 
 extracted using standard regions (the source cell encompassing 
97\% of the cross-dispersed counts) and extraction criteria, 
resulting in exposure times $\sim 114$ ks in RGS-1 (R1) and 
$\sim 108$ ks in RGS-2 (R2).  
The full-band (0.34-2.0 keV) RGS background-subtracted count rates were 
$0.062\pm0.0008$ (R1) and $0.067\pm0.0008$ (R2). 
The background comprised 17\% of the total count rate in 
R1 and R2, respectively.  We note that due to problems with 
some of the CCD chips onto which the RGS spectrum is  dispersed there are 
two prominent data gaps evident, R1 has a gap between $\sim$ 0.9-1.2 keV 
and R2 has a gap between $\sim 0.51$ - 0.62 keV. 
We note that in this analysis we adopted cz=2649 km/s \citet{keel96}. 

Ignoring emission and absorption lines 
for the moment, we find a good parameterization of the   
continuum shape in RGS data over the 0.3 - 2 keV band 
 using a powerlaw of photon index $\Gamma=2.17^{+0.04}_{-0.06}$ 
 attenuated 
by a Galactic column of cold gas $N_H=2.9 \times 10^{20} {\rm cm^{-2}}$ 
plus a column of ionized gas with
 $N_{H(UV)} = 7.7 \times 10^{21}  {\rm cm^{-2}}$. 
Such a column of ionized gas 
is expected from the UV and LETG analyses \citep{netzer02,kraemer02} 
of data from an epoch close in time to these RGS data (and the ionized 
gas is modeled using CLOUDY90 for solar abundance material, \citealt{f98}). 
Inclusion of 
the ionized gas improves the fit from $\chi^2=1594/639\ dof$ to 
$1180/638\ dof$ and provides a good model to the gross shape of the 
soft spectrum.  Figure~2 shows 
the data compared to this model. 
However, our parameterization 
of the continuum shape is not the only possible model. 
A recent alternative applied to some Seyfert spectra   
parameterizes broad spectral wiggles as emission lines due 
to ionized material in the accretion disk (e.g. \citealt{br}). However, we do 
not pursue 
such a parameterization in this paper.  The  
 absorption edges expected from the (previously observed) 
ionized absorber provide a good description of most of the broad features,  
and we believe this makes our interpretation of the spectrum a  
compelling way to proceed for this observation of NGC~3516.

The most immediate result evident from Figure~2 is that 
the RGS spectra show several prominent emission lines. 
The two strongest lines 
are those at 0.4198 keV and 0.5614 keV. We identify these with 
forbidden components of N{\sc vi} and O{\sc vii}, respectively. These 
two strong lines were measured carefully to determine 
bulk and turbulent velocities of the emitting gas. 
For the N{\sc vi} line we measured a rest-energy 
$E=419.8\pm0.2$ eV and for the O{\sc vii} line 
$E=561.4\pm0.3$ eV. Errors are 90\% confidence. 
Intrinsic line widths were found to be 
$\sigma=3.8^{+3.5}_{-3.8} \times 10^{-4}$ keV 
and   $\sigma=5.6^{+4.4}_{-5.6} \times 10^{-4}$ keV, respectively.  
The energy for O{\sc vii} {\it f} corresponds to blueshift 
of $214^{+160}_{-80}$ km/s while for N{\sc vi} the blueshift is 
$143^{+143}_{-143}$ km/s, i.e. consistent with no blueshift. 
 For comparison, the
absolute wavelength calibration for the RGS is $\sim 8$m \AA,
or $\sim 100$ km/s at 0.5610 keV. Thus the evidence for a blueshift 
for the O{\sc vii} line is marginal, the lower limit 
is close to the accuracy of the detector absolute 
wavelength calibration. 
The line widths correspond to velocities 
FWHM(N{\sc vi}) $=646^{+581}_{-646}$ km/s and FWHM(O{\sc vii}) 
$=704^{+554}_{-704}$ km/s. Line widths are consistent with zero, 
the upper limits 
constrain velocity broadening to $\lesssim 1300$ km/s. 
(For comparison the FWHM resolution 
of R1 is $\sim 680$ km/s at the observed energy of N{\sc vi}, 0.416 keV 
and $\sim 850$ km/s at the observed energy of O{\sc vii}, 0.556 keV). 
Changes in the detector effective area with energy cause some apparent
asymmetries at first glance as the 
effective area changes significantly across some line profiles 
(e.g. O{\sc viii} at 0.654 keV). However we find no true 
asymmetry in any emission line profile. 

To search for weak emission lines we slid a Gaussian 
template across the data, testing for an improvement to the fit at 
every resolution element compared to our
model of the underlying continuum. 
The data were first binned by a factor of 
8 to a binsize of 0.32 \AA\ to ensure that there were greater than 
20 photons in each spectral bin so $\chi^2$ fitting could be employed. 
Examination of the RGS background spectra shows 
a broad bump centered around 0.39 keV (not 
fully understood but thought 
likely due to relatively high dark current in CCD2 for both RGS instruments; 
XMM Users Handbook, also see \citealt{denh}). Other known 
features in the  RGS background include 
emission from Al K$\alpha$ close to 1.5 keV, due to Al in the  
detector housing. 
This means we need to subtract the RGS background 
to ensure we isolate the spectral properties of the source. 
Figure~3 shows the results of this sliding Gaussian 
test against the observed data in red.
Plotted is the improvement in the fit-statistic. 
This the plot is useful in assessing the
{\it significance} of each line, but not the strength of the line. 
It can be seen that the addition of an emission feature results 
in a large improvement in $\chi^2$-statistic at several energies.

In order to assess the reality of these features, 
we have constructed a series of simulated datasets. 
For each RGS, 20 separate background spectra were 
generated by randomly adding the appropriate 
Poisson noise to each channel in the observed background spectrum.
Similarly, 20 simulated source spectra were generated for each
RGS, again with Poisson noise introduced, using our 
model of the underlying continuum convolved with the 
instrumental response. 
These simulated data sets were then combined
and 20 R1+R2 pairs analysed in the same manner as  
the observed spectra above. Specifically, the sliding Gaussian 
test was performed for each of the 20 pairs, and the maximum 
reduction in fit-statistic at each energy noted. The result is 
plotted in blue in Figure~3, and therefore represents a 
95\% significance threshold at each energy (and takes 
into account all instrumental and background features).

Including the N{\sc vi} and O{\sc vii} lines discussed above, 
we find a total of 13 emission features (some of which are blends)
in the observed spectrum above the 95\% confidence threshold.
Ten of these 13 occur at energies consistent with atomic transitions 
of abundant ions, and have understandable intensity ratios
(below). 
Thus we consider these 10 feature to be robust detections.
They are labeled on Figure~3 and listed in Table~1. 
Of the remaining three features, that at 0.39~keV is most likely 
due to the detector background described above.
The final two features (at 0.697 and 0.706~keV; and 
apparent strengths corresponding to 
$\sim 0.4 \times 10^{-5}$ photons cm$^{-2}$ s$^{-1}$)
do occur at energies close to atomic transitions. 
The former, is consistent with the O{\sc vii}
($1s^2 \rightarrow 1s4p$) resonance line at 0.698 keV.
However, the lack of any indication of the O{\sc vii}
($1s^2 \rightarrow 1s3p$) line at 0.665~keV makes such an 
interpretation extremely unlikely. 
The line at 0.706~keV could be identified with S{\sc xiv} RRC at
0.707~keV.  However, such an interpretation would lead us to expect a 
much stronger O{\sc viii} RRC at 0.871~keV than observed.  
In summary, only the 10 emission lines which we believe 
secure ($\delta \chi^2 > 10$) and attributable to NGC~3516 appear in Table~1. 

While some of the detected features have some significant width, e.g. the
C{\sc vi} RRC, the features were too weak for us to
fit for the width and 
thus the temperature of the
emitting gas. However we did obtain an 
upper limit on width, $kT \lesssim 0.01$keV,
i.e. $T \lesssim 120,000 $K. This is consistent with the width of the
RRC features in NGC 1068, based on which \citet{kinkhabwalaea02} argued
that the X-ray emitting plasma was purely photoionized. 

The emission lines were assumed to originate outside of the UV absorber 
(i.e. not absorbed by it). If the emitter were inside the UV absorber it would 
have to reside within $10^{16}$ cm of the nucleus with a density 
$10^9$ cm$^{-3}$ or so \citep{kraemer02}. While such a location is possible, 
the velocity widths of the emission lines at FWHM$< 1300$ km/s are 
rather narrow for such a small radial location. These widths are 
more suggestive 
of an origin on the pc-scale. Furthermore, some results from 
Seyfert~2 galaxies suggest the X-ray emitting gas in general 
has extent on the 10-100pc size-scale (e.g. \citealt{sako00}).   
All things considered, it seems most likely the emitting gas lies outside of 
the UV absorber in this case. The fact the absorption edges are 
imprinted on the soft spectrum must then be explained as the 
soft-band having a significant contribution from nuclear continuum 
emission, transmitted through the UV absorber (the fraction of the nuclear
continuum scattered by the line-emitting gas is negligible, as shown in Section 3.2).
Thus we suggest a full model 
of absorbed continuum component plus unabsorbed lines (with 
Galactic column absorbing all).  The line fluxes in Table~1 
are the inferred intrinsic 
fluxes at the source (i.e. corrected for Galactic absorption). 
In addition to tabulating the fitted line fluxes, 
we noted the total counts in the bin, those attributable to the 
 continuum emission and thus those attributable to the emission line itself. 
Errors are calculated on the counts in the line and the same fractional 
errors are applicable to the fluxes listed. 

The process was repeated to search for weak absorption lines.
Following a similar process to that used to detect
the emission lines, we found evidence for an absorption feature
at 0.358 keV and  broad features centered on 0.753 keV and 0.96 keV.
The former
has no obvious identification. The feature at 0.75 keV, while close to a jump
in detector efficiency, appears real, and we identify this with
the unresolved transition array due to inner-shell absorption by
Fe{\sc ix - xi} M-shell ions, as observed in some
other AGN (e.g. IRAS 13349$+$2438, \citealt{sakoiras01}).  The feature
at $\sim 0.96$ keV may be a blend of Fe lines from highly-ionized gas
(Fe {\sc xx} - {\sc xxiv}). Much weaker features at 0.54 keV and 0.61 keV
do not have clear identifications.

Absorption lines and edges were  expected to be detected in NGC~3516 from
transmission of the nuclear continuum through ionized gas, previously
observed in this source (e.g., \citealt{kriss96,netzer02,kraemer02}).
The overall shape of
the RGS spectra shows the  imprint of the absorption edges
expected from the X-ray absorption associated with the gas measured
by STIS \citep{kraemer02}. The width of the UV absorption lines is
$\sim 20 - 70 $ km/s,  X-ray absorption lines with the same velocity
structure would be too narrow to detect with the RGS (or
{\it Chandra} grating data).

\section{FUSE data}

We reduced data from an archival 
{\it Far Ultraviolet 
Spectroscopic Explorer (FUSE)} spectrum, originally obtained
on 2000 April 17, again,  while NGC 3516 was in an extremely low flux state
(see \citealt{kraemer02}). From the {\it FUSE} spectrum 
we measured an O{\sc vi} $\lambda$ 1038 flux of 
$\approx 1.6 \times 10^{-14}$ ergs s$^{-1}$ cm$^{-2}$, 
with a FWHM of $\sim$ 250 km s$^{-1}$. 
Given the intrinsic 2:1 
emission ratio of O{\sc vi} $\lambda$ 1032/ O{\sc vi} $\lambda$ 1038,
line fluxes 
the total O{\sc vi} flux should be $\sim$ 4.0 x 10$^{-14}$ ergs s$^{-1}$ 
cm$^{-2}$. However, as noted by \citet{hutchings01} in their analysis
of this spectrum, the observed O{\sc vi} doublet ratio is $\sim$ 1:1, 
which they attributed 
to line transfer effects in optically thin gas. We think that it is more
likely that the lines are absorbed by intervening gas, which may explain
the asymmetric line profiles (see \citealt{hutchings01}) and the narrowness
of the O{\sc vi} lines compared to the X-ray lines. However, it is not
necessary that this absorption arises in 
the same material that produces the deep, variable X-ray absorption discussed
in \citet{netzer02} and \citet{kraemer02}, but could be a relatively
small column of ionized gas (e.g., $\sim$ 10$^{20}$ cm$^{-2}$, $U$ $\sim$ 0.5)
that covers the NLR of NGC 3516. 
This component may only be detectable when viewed
against the NLR emission, much of which was missed in the 
small 0\arcsecpoint2 $\times$
0\arcsecpoint2 STIS aperture, but is 
well-covered by the 20\arcsecpoint0 $\times$ 20\arcsecpoint0
{\it FUSE} aperture. (At a distance of 
35.7 Mpc for NGC~3516, $1'' = 173$pc.)  
In fact, similar evidence for absorbed
UV narrow-line profiles has been seen in STIS spectra of NGC 1068 
\citep{kraemerc00}.

\section{Physical Conditions in the Emission-Line Gas}

\subsection{Emission-Line Diagnostics}

The dominance of X-ray emission lines in the soft spectrum of 
this Seyfert 1 galaxy is consistent with the 
low flux state of the source. If these lines have a 
$\sim$constant flux over years, then they would have been swamped 
by the nuclear emission at higher flux states. 

This source is one of the few observed to date 
with detectable  forbidden emission lines from 3 abundant
elements in the He-like isoelectronic sequence (N{\sc vi}, O{\sc vii}
and Ne{\sc ix}). It is the first such case for a Seyfert 1 galaxy. 
Even at first glance, the emission line ratios look unusual. 
 Compare for example the {\it XMM} RGS spectrum of NGC~3783 where the 
N{\sc vi} line was very weak compared to O{\sc vii}, as expected 
for solar abundance gas \citep{blustinea02}. The details of these
anomalous line ratios are discussed, in the context of the
photoionization models, in the following section. 

As suggested by various authors (\citealt{pradhana85,liedahld99,
mewer99,porquetd00}), the physical conditions of 
photoionized plasmas can be constrained via the relative strengths of the most 
intense lines from He-like species; the resonance line (1s$^{2}\ $$^{1}$S$_{0}$
-- 1s2p\ $^{1}$P$_{1}$), the two 
intercombination lines (1s$^{2}$\ $^{1}$S$_{0}$ -- 1s2p\ $^{3}$P$_{2,1}$), and
the forbidden line (1s$^{2}$\ $^{1}$S$_{0}$ -- 1s2s\ $^{3}$S$_{1}$). The ratios
of the forbidden line to the two intercombination lines (the $R$ ratio) is 
sensitive to
density, while the ratio of the sum of the forbidden and intercombination
lines to the resonance line (the $G$ ratio) is sensitive to electron 
temperature. If the
resonance lines are relatively strong, the latter ratio can indicate
temperatures in excess of those expected for a photoionized plasma. However, 
the resonance lines can be also be photo-excited \citep{sako00}, hence 
the $G$-ratio may not be a reliable indicator of either the temperature or means
of excitation of the plasma. Based on their analysis of an {\it XMM-Newton}
RGS spectrum of NGC 1068, \citet{kinkhabwalaea02} suggested that
the excess emission in virtually all the detected resonance lines was the
result of photo-excitation in a purely photoionized plasma. 

In these data, the $R$ ratio of the O{\sc vii} f-line 
to the i-lines is $\sim$ 3.5 (see Table 1), which
for a photoionized plasma implies an electron density n$_{e}$ $\sim$ several
x 10$^{9}$ cm$^{-3}$ \citep{porquetd00}. However, the weakness
of the intercombination lines, plus any underlying contribution from the wings
of the adjacent forbidden and resonance lines, can easily lead to an
overestimate of the flux. Hence, it is likely that n$_{e}$ is lower.
Furthermore, at sufficiently high density, the metastable 
1s2s\ $^{3}$S$_{1}$ level begins to collisionally depopulate, suppressing the
forbidden line of the He-like triplet. For N{\sc vi}  
this occurs at n$_{e}$ $>$ 10$^{8}$ cm$^{-3}$, while the depopulation
of corresponding O{\sc vii}  level occurs at densities $\sim$ factor of 10 greater
\citep{porquetd00}. Given the relative strengths of the N{\sc vi} and
 O{\sc vii}  
forbidden lines, we suggest that the average density in the 
emission-line gas is n$_{e}$ $\lesssim$ 10$^{8}$ cm$^{-3}$.

\subsection{Photoionization Models}

Photoionization models of the X-ray emission-line gas were generated using the
Beta 5 version of CLOUDY (Gary Ferland, private communication). 
This version of 
the code includes new atomic data for the He-like ions. We modeled the 
emission-line gas as a single-zoned slab of
atomic gas, irradiated by the central source, for which we used the
same spectral energy distribution and luminosity 
($\sim 10^{43}$ erg\ s$^{-1}$) described in \citet{kraemer02}.
As usual, the models are parameterized in terms of
the ionization parameter $U$, which is the ratio of ionizing photons
per H atom at the ionized face of the slab. We {\it initially} assumed 
solar elemental abundances \citep{ga89},
which are, by number relative to H, as follows: He $=$ 0.1,
C $=$ 3.4 x 10$^{-4}$, N $=$ 1.2 x 10$^{-4}$, O $=$ 6.8 x 10$^{-4}$,
Ne $=$ 1.1 x 10$^{-4}$, Mg $=$ 3.3 x 10$^{-5}$, Si $=$ 3.1 x 10$^{-5}$,
S $=$ 1.5 x 10$^{-5}$, and Fe $=$ 4.0 x 10$^{-5}$. The gas was assumed
to be free of dust. 

Prior to generating new models, we explored the possibility that the
X-ray emission lines arise in the UV/X-ray absorber 
described in \citet{kraemer02}. 
In the current low-flux state of NGC 3516, the models
of the strongest UV components predicted ratios of ionic columns for 
 O{\sc vi}:O{\sc vii}:O{\sc viii} to be  
 roughly 1.00:0.67:0.03. Consequently, little  O{\sc viii} Ly$\alpha$ emission
is expected relative to O{\sc vii} emission for  
the UV gas seen in emission. This is not the case, we observe  
enough O{\sc viii} emission to know there must be an additional  
 component of higher-ionization gas present. 
We also note 
the model for the production of 
O{\sc vi} $\lambda\lambda$ 1032,1038 emission lines from the UV gas 
 yields fluxes $\sim$ 20 times as strong as the O{\sc vii} f-line. 
In the data they appear 
weaker than this, as noted further below. 

Since the X-ray emission-lines cannot be formed in the UV/X-ray absorbers,
we permitted $U$ and the total column density 
(N$_{H}$ $=$ H{\sc i} $+$ H{\sc ii}) to vary, as a new model is required for 
the X-ray emitting gas.  
In Table 2, we compare the model predictions to the measured emission-line 
fluxes by scaling the predicted O{\sc vii} f-line flux to that
observed. We find that line ratios can be
roughly matched using a single-zoned model, with $U$ $=$ 1.1, N$_{H}$ $=$ 
1.5 x 10$^{21}$ cm$^{-2}$, and solar abundances (Model 1).
In order to boost the strength of the resonance lines via photo-excitation, 
we introduced
a turbulent velocity of 50 km s$^{-1}$, which is roughly equal to the
average FHWM of the
individual UV absorbers \citep{kraemer02}.
If we assume n$_{e}$ $=$ 10$^{8}$ cm$^{-3}$, our model results place the
emission-line gas at a radial distance of 0.02 pc; although this is
within the range determined for the UV absorbers, since the 
lower limit to the density 
is not well constrained. It is quite plausible that
the emission-line gas lies further from the nucleus, as we discussed
in Section 2. The model
predicts a mean electron temperature of 7.4 x10$^{4}$K, in agreement with the
constraints determined from the C{\sc vi} RRC.

The model predictions assuming solar abundance 
are within the measurement errors 
for the O{\sc vii} He-like lines, Ne{\sc ix} {\it f}, and O{\sc viii} 
Ly$\alpha$, 
while the prediction for C{\sc vi}  Ly$\alpha$ is slightly
high. However, the strengths
of the N{\sc vi} He-like lines and N{\sc vii} Ly$\alpha$ line are
unpredicted by factors of $\sim$ 4 and $\sim$ 2, respectively, which is
strong evidence against our initial assumption of solar abundances.
The apparently anomalous line ratios in NGC~3516 suggests one of two 
possibilities; the other elements such as C and O are 
depleted onto dust grains (e.g. \citealt{shieldsk95}), or the N 
abundance exceeds the Solar value. 
 
Considering the case of depletion onto dust grains. 
UV data show many cases of enhanced nitrogen relative to carbon, e.g. 
 NGC 5548 \citep{crenshaw03} and Akn 564 
\citep{crenshaw01}. In those cases C may be depleted onto 
dust grains. However, in NGC~3516 the 
 data argue against dust depletion. While the ratio of  N/C  is high, 
the ratio N/O is also high and yet 
O is not heavily depleted onto dust grains (see \citealt{snoww96}). 
Also the ratio N/Ne is high and 
Ne cannot be depleted onto dust grains. Thus we rule out depletion.
 
Hence, it is most likely that nitrogen is over-abundant, perhaps
by a factor of 2 -- 3. Interestingly, \citealt{kinkhabwalaea02} came to a 
similar conclusion for NGC 1068. As a result, we regenerated the model using
a N/H ratio of 2.5 x solar (Model 2). As shown in Table 2, the fit for the
N{\sc vii} line is quite good. 
While the N{\sc vi} lines are much closer to the
observed strengths, they are still somewhat 
underpredicted. The latter may simply be a reflection of the difficulty in 
matching the emission lines with a single-zone model, since some additional
N{\sc vi} could arise in a component of lower ionization gas.  
Our discovery of nitrogen overabundance led us to revisit the absorber 
used which fits the UV data and models the broad features in the 
soft X-ray band. First we note that 
the STIS data on which that absorber was based 
\citep{kraemer02} showed saturated N and C lines, 
so limits from those lines 
 did not provide tight constraints on the state of the gas for the UV model. 
Fine-tuning  
abundances for the absorber model does not make any significant difference 
to the fit to broad features and thus to the 
emission line analysis and conclusions drawn here. 

Based on the model predictions, we can constrain the global covering factor
of the emission-line gas. Our predicted flux for O{\sc vii} {\it f} emitted
from the illuminated face of the photoionized slab is 
1.7 x 10$^{5}$ ergs cm$^{-2}$ s$^{-1}$.
Adopting a distance of 
35.7 Mpc for NGC~3516, assuming H$_{o}$ $=$ 75 km s$^{-1}$ Mpc $^{-1}$  
\citep{fwm98}, the total O~VII {\it f} emission
from NGC 3516
is $\approx$ 5.5 x 10$^{39}$ ergs s$^{-1}$. Hence, the emitting surface
area must be $\approx$ 3.2 x 10$^{34}$ cm$^{2}$. Based
on our assumed density, n$_{e} $=$ 10^{8}$ cm$^{-3}$, and luminosity in ionizing
photons, the emission-line gas lies at a radial distance of 6.9 x 10$^{16}$ cm,
which yields a total surface area of 6.0 x 10$^{34}$ cm$^{2}$. 
Hence the covering factor of the
emission line gas is $F_{c}$ $\sim$ 0.5. While we expect that the emission-line
gas extends over a range in radial distance, and hence density, the total
covering factor will be similar.

From the predicted covering factor and column density, we can determine the
fraction of the nuclear continuum reflected via electron scattering within the
emission-line gas. Assuming isotropic scattering, at small electron-scattering optical depths 
$\tau_{e}$  $<$ 1, the reflected fraction of continuum radiation $f_{r}$ $\approx$
N$_{electron}$ $F_{c}$ $\sigma_{T}$, where N$_{electron}$ is the electron
column density ($\approx$ N$_{H}$) and $\sigma_{T}$ is the Thomson cross-section.
We find $f_{r}$ $\approx$ 5 x 10$^{-4}$, hence a negligible fraction of continuum
radiation will be scattered into our line-of-sight by this component. 

Returning to the question of the strength of 
O{\sc vi} emission. 
The total O{\sc vi} flux derived from the {\it FUSE} data 
is $\sim$ 4.0 x 10$^{-14}$ ergs s$^{-1}$ cm$^{-2}$. 
Our X-ray emission-line model predicts a total O{\sc vi} flux
of 6.75 x 10$^{-14}$ ergs s$^{-1}$ cm$^{-2}$, which, since 
it is somewhat higher 
than the {\it FUSE} value, may support the indication 
(from the doublet ratio) that these lines 
are absorbed. Under the assumption that the strengths of the emission 
lines are constant this O{\sc vi} measurement 
 provides a tight constraint on the ionization state of the 
NLR gas in which the UV and optical emission-lines arise.

\section{Discussion}

The relation between abundance and 
AGN redshift, luminosity, and the inter-relation between the AGN and 
starburst regions are key to understanding AGN formation and evolution. 
Based on the ratios of the N{\sc v} $\lambda$ 1240 emission line to C{\sc iv} 
$\lambda$ 1550 and
He{\sc ii} $\lambda$1640, it has been argued that 
QSOs show high metallicity (Z$\geq$
1, where Z $=$ 1 for solar abundances) and,
in particular, high N abundance at 
redshifts as high as $z$ $\geq$ 3, indicative of 
enrichment due 
to rapid star formation at epochs as early as $\leq$ 1 Gyr \citep{hamann93}.
Among relatively nearby ($z$ $\lesssim$ 0.05) AGN, Narrow-Line Seyfert 1s show evidence of large N 
abundances (e.g., \citealt{wills99}). It is, therefore, interesting that the X-ray spectra of 
NGC 3516, a low-redshift,
Broad-Line Seyfert 1 galaxy, also shows evidence for an overabundance by a factor
of 2 --3 of nitrogen in the central regions of the galaxy, 
with respect to other heavy elements. Furthermore,
\citet{kinkhabwalaea02} presented evidence for a similar enhancement of nitrogen in the 
prototypical Seyfert 2 galaxy NGC 1068. Also, based on the strength of the N~V $\lambda$1240
line, there is evidence of anomalous
N abundances in the NLRs of the Seyfert 1 galaxies 
NGC 5548 \citep{kraemer98} and NGC 4151 \citep{kraemer00}, although this could be at least
partly due to enhanced emission by photo-excitation. 

Nitrogen can be synthesized from carbon and oxygen produced 
within a star, referred to 
as primary production, or via the CNO process in intermediate mass
stars (M $\lesssim$ 7 M$_\odot$; see \citealt{maeder89}) which already possess
C and O, i.e., secondary production (e.g., \citealt{tinsley80}).
At certain temperatures N is enhanced at the expense of C (while at higher 
temperatures it would be at the expense of O). Thus it is possible to observe 
solar O, Ne etc yet obtain enhanced N with subsolar C. 
The enrichment of nitrogen via secondary production increases with 
overall metallicity such that N/H $\propto$ (O/H)$^{2}$ $\propto$ Z$^{2}$. 
Since the evidence for overabundant nitrogen is typically the high ratios of
N lines to those of C or other heavy elements, one might expect that
a factor of 3 enhancement of N/O would require Z $\sim$ 3 and N/H
$\sim 9$ times solar! However, for Z $\sim$ 1, it is still possible to have 
N/H a few times solar, assuming the enhancement of nitrogen is accompanied
by a loss of carbon and, possibly oxygen (e.g. \citealt{maeder89}). For 
example, assuming roughly
solar initial abundances \citep{ga89}, N/H could be 2.5 times solar
if approximately half of the carbon were used up in the production
of nitrogen. Interestingly, our models with solar carbon abundance overpredicted the strength of the
C{\sc vi} L$\alpha$ line by a factor of $\sim$ 1.5.
To explore the possibility that some of the carbon was lost in the production
of nitrogen, we generated a third model, with the carbon abundance set to 47\% solar (to fully
account for the enhancement of nitrogen) and all other model parameters
kept identical to those of Model 2. The predicted line fluxes for Model 3, scaled to the
O{\sc vii} f-line, are listed in Table 2.
The C{\sc vi} L$\alpha$ line is now within the measurement errors, while
the predictions for the other lines are essentially unchanged from those of Model 2.
Hence, the emission-line fluxes are consistent with emission from photo-ionized
gas in which the nitrogen is enhanced and the carbon depleted by the same
amount. If the anomalous N/O and N/H ratios are consistent
with Z $\sim$ 1 and are the result of nitrogen production in 
intermediate mass stars, these results provide tight constraints
on the history of star formation in the nucleus of NGC 3516.   

\section{Conclusions}

We have used {\it XMM-Newton} RGS grating spectral data 
to examine the physical conditions within the
X-ray emission-line gas in the Seyfert 1 galaxy NGC 3516.
The spectra show emission-lines from the He-like ions of N,
O and Ne, and H-like ions of C, N, and O. Also, we have detected
RRC from C{\sc vi} and O{\sc viii}. We have shown the following:

1. The RGS data show the soft X-ray absorber 
to be consistent with the UV absorbers detected in earlier
{\it HST}/STIS observations. However, the UV absorbers
cannot account for the X-ray line emission. 
 Although 
the UV absorbers could account for some of the N{\sc vi} and O{\sc vii} 
emission, they are generally too low an ionization state
to produce any of the higher ionization lines detected in the RGS
spectra. Furthermore, if the emission from the UV absorbers were
scaled to fit the O{\sc vii} {\it f} component, the models predict O{\sc vi}  
$\lambda\lambda$ 1032,1038 lines almost an order of
magnitude stronger than observed in a recent {\it FUSE} spectrum. 
While some line variability may occur, an order of magnitude change in 
O{\sc vi} between the {\it FUSE} observation and the epoch reported here 
is highly unlikely. 

2. From the C{\sc vi} RRC, we find $kT$ $\lesssim$ 0.01 keV, consistent
with low temperature ($\lesssim$ 10$^{5}$ K), photoionized gas.  
Based on the ratios of the f- and i- lines from He-like
O{\sc vii} and the relative strength of N{\sc vi} {\it f}, we suggest that
the gas is in the low-density regime (n$_{e}$ $\lesssim$ 10$^{8}$ cm$^{-2}$).
We have been able to fit the observed emission-line ratios
with a single zoned photoionization model, with N$_{H}$ $=$
1.5 x 10$^{21}$ cm$^{-2}$ and $U$ = 1.1, however, in the case
of solar abundances, the N{\sc vi} and N{\sc vii}  
lines are significantly underpredicted. 
Hence, we suggest that the N/H ratio is at least 2.5 times solar, which
may be the result of secondary production of nitrogen in intermediate
mass stars. Follow-up observations of the stellar population
in the nucleus of NGC 3516 could help test this possibility.  

3. In order to produce the observed O{\sc vii} line fluxes, the emission-line
gas must have a global covering factor of $\sim$ 0.5. The scaled (predicted) 
O{\sc vi} $\lambda\lambda$ 1032, 1038 emission is slightly higher than that
seen in the {\it FUSE} spectrum, although the O{\sc vi} line ratios and
profiles show strong evidence for absorption, possibly from an additional
component of UV absorption near systemic that cannot be deconvolved from the
strong UV components seen in the STIS spectra. Hence, the X-ray model
is generally consistent with the {\it FUSE} spectrum.

\section{Acknowledgements}

We are grateful to the  {\it XMM}   
satellite operation team and to Martin Still for help with the RGS data.   
T.J.\ Turner acknowledges support from NASA 
grant  NAG5-7538. We thank Tahir Yaqoob, 
Fred Hamann, Allen Sweigart, Fred Bruhweiler, 
Mike Crenshaw, Jose Ruiz and the anonymous referee 
for useful comments. We especially
thank Gary Ferland and Ryan Porter for their 
generosity in allowing us to use the latest pre-release version of CLOUDY.

\clearpage

\begin{deluxetable}{lcccc}
\tablewidth{0pc}
 \tablecaption{RGS Emission Line Measurements\label{rgs}}
 \tablehead{
 \colhead{Energy\tablenotemark{a}} &
 \colhead{ID} &
 \colhead{Line\tablenotemark{b} } &
\colhead{Flux1}\tablenotemark{c} &
\colhead{Flux2}\tablenotemark{d}}
 \startdata

0.3671 & C{\sc vi}Ly$\alpha$ (0.366) & $46\pm15$ & 1.85 & 1.08\\
0.4198 & N{\sc vi}{\it f} (0.4198)   &$119\pm17$ & 3.18 & 2.13 \\
0.4280 & N{\sc vi} {\it i$+$r} (0.4263/0.4307) & $41\pm13$ & 1.69 & 1.16\\
0.4913 & C{\sc vi} RRC (0.4900)  & $29^{+10}_{-11}$  & 3.35 & 2.63\\
0.5004 & N{\sc vii} (0.5002)   & $37\pm14$ & 1.06 & 0.85 \\
0.5614 & O{\sc vii}(f) (0.5610) \tablenotemark{e}   & $122\pm17$ & 4.21 & 3.78\\
0.5681 & O{\sc vii}(i) (0.5686) \tablenotemark{e}  & $30\pm12$ & 1.20 & 1.09\\
0.5727 & O{\sc vii}(r) (0.5739) \tablenotemark{e} & $49\pm12$ & 1.38 & 1.27\\
0.6540 & O{\sc viii} (0.6510)  & 48$\pm12$ & 0.82 & 1.31 \\
0.9047 & Ne{\sc ix} {\it f} (0.9055) \tablenotemark{f}
	 & $25^{+10}_{-9}$ & 0.57 & 0.83 \\
		\enddata
\tablenotetext{a}{The measured rest-energy of the line in keV, 
corrected for the systematic velocity of the host galaxy}  
\tablenotetext{b}{Counts attributed to the line (after subtraction of the continuum flux). Errors are 1 $\sigma$}
\tablenotetext{c}{Total photons cm$^{-2}$ s$^{-1}$ in the line in units
$10^{-5}$. Lines assumed to have widths FWHM$=650$ km/s}
\tablenotetext{d}{Line flux as ergs cm$^{-2}$ s$^{-1}$ in the line in units
$10^{-14}$} 
\tablenotetext{e}{RGS 1 only}
\tablenotetext{f}{RGS 2 only}
\end{deluxetable}

\begin{deluxetable}{lcccc}
\tablewidth{0pc}
 \tablecaption{Predicted Emission-Line Fluxes\tablenotemark{a}}
 \tablehead{
 \colhead{Line} &
 \colhead{Model 1\tablenotemark{b}} &
\colhead{Model 2\tablenotemark{c}} &
\colhead{Model 3\tablenotemark{d}} &
\colhead{Measured}}
 \startdata
C{\sc vi}Ly$\alpha$ & 1.55 & 1.57 & 0.87 & 1.08 $\pm$ 0.35 \\ 
N{\sc vi}{\it f} & 0.57 & 1.43 & 1.44 & 2.13 $\pm$ 0.30\\  
N{\sc vi} {\it i$+$r} &  0.48 & 0.86 & 0.86 &  1.16 $\pm$ 0.37 \\
N{\sc vii} & 0.53 & 1.10 & 1.19 & 0.85 $\pm$ 0.32 \\
O{\sc vii}(f)  & 3.78 & 3.78 & 3.78 & 3.78 $\pm$ 0.53\\
O{\sc vii}(i) & 0.94 & 0.94 & 0.94 & 1.09  $\pm$ 0.44 \\
O{\sc vii}(r) & 1.03 & 1.03 & 1.04 & 1.27 $\pm$ 0.31 \\ 
O{\sc viii} & 1.30 & 1.27 & 1.30 & 1.31 $\pm$ 0.33 \\
Ne{\sc ix} {\it f} & 0.89 & 0.90 & 0.89 & 0.83 $\pm$ 0.33 \\
		\enddata
\tablenotetext{a}{Fluxes as ergs cm$^{-2}$ s$^{-1}$ in units of 10$^{-14}$.}
\tablenotetext{b}{Solar abundances. Fluxes scaled to the O~VII f-line.}
\tablenotetext{c}{N/H $\sim 2.5$ times solar. Fluxes scaled to the O~VII f-line.}
\tablenotetext{d}{N/H $\sim 2.5$ times solar; C depleted to 47\% relative to solar. Fluxes scaled to the O~VII f-line.}
\end{deluxetable}

\clearpage

\typeout{FIGS}

\begin{figure}   
         \epsscale{0.7} 
       \plotone{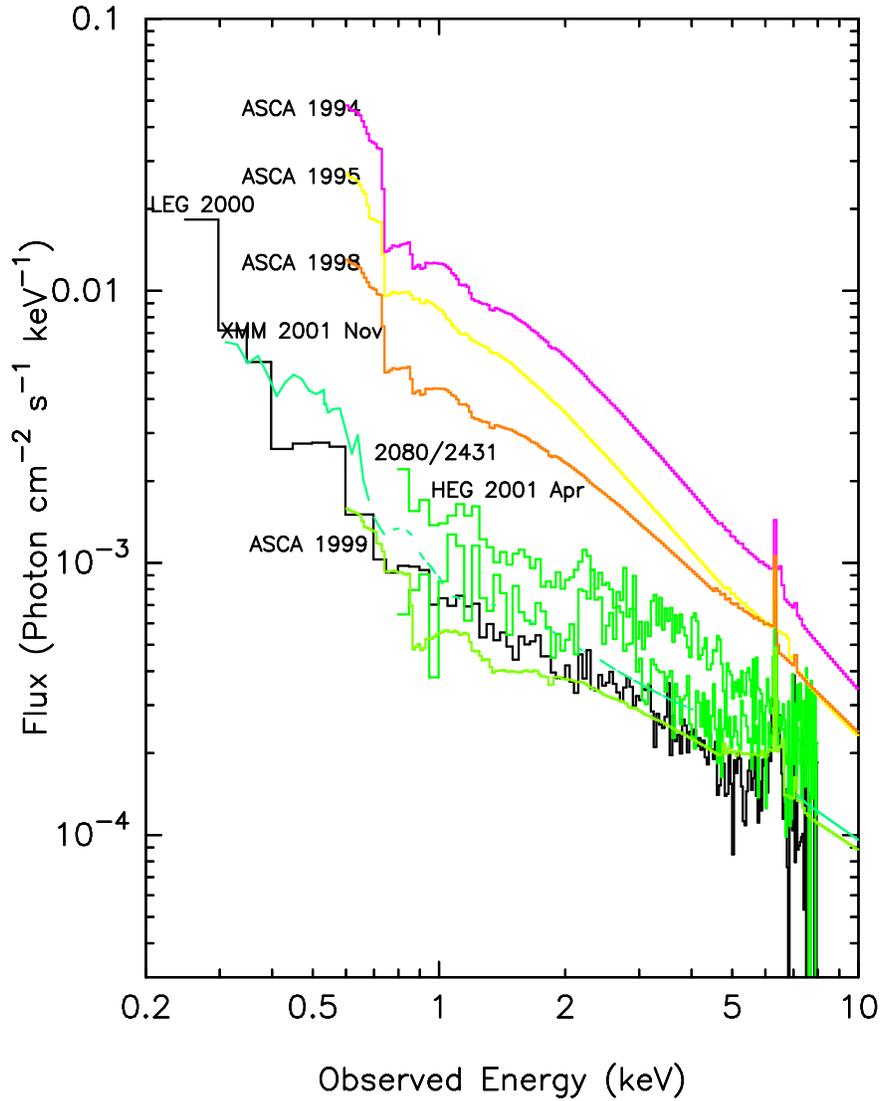}         
	        \caption[Flux Levels]          
{An X-ray flux history for NGC 3516. Schematic models are shown 
representing the X-ray flux levels observed by recent satellites. 
The data show NGC~3516 to be close to the 
lowest flux state on record, also 
observed by the LETG and by {\it ASCA} in 1999. 
 HEG spectra from two observation intervals 
2001 April are represented, as are lines showing the 
 source at 
a much higher flux state, as observed by {\it ASCA} }
\end{figure}

\begin{figure}

\plottwo{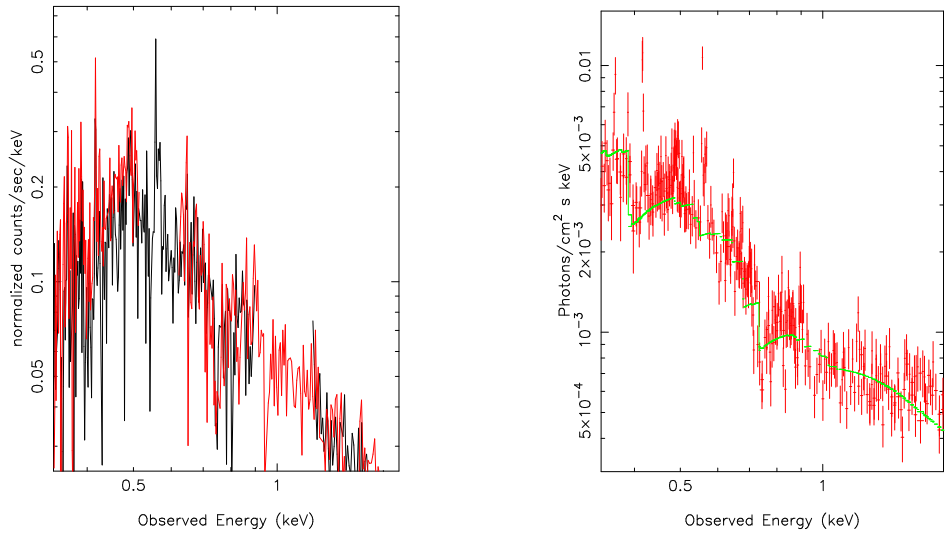}{f2b.eps}

\caption{Left: The RGS data with errors removed for clarity of viewing. 
The red line represents R2, the black line represents R1. The 
data points are connected by a line, the gaps evident in 
one section of each RGS are due to loss of functionality in some 
 CCD chips. 
Right: The unfolded RGS 1 and 2 data (red) compared to a 
powerlaw model attenuated by 
the absorbing column indicated by UV data (green line)} 
\label{fig:spec}
\end{figure}

\begin{figure}
\includegraphics[scale=0.7,angle=270]{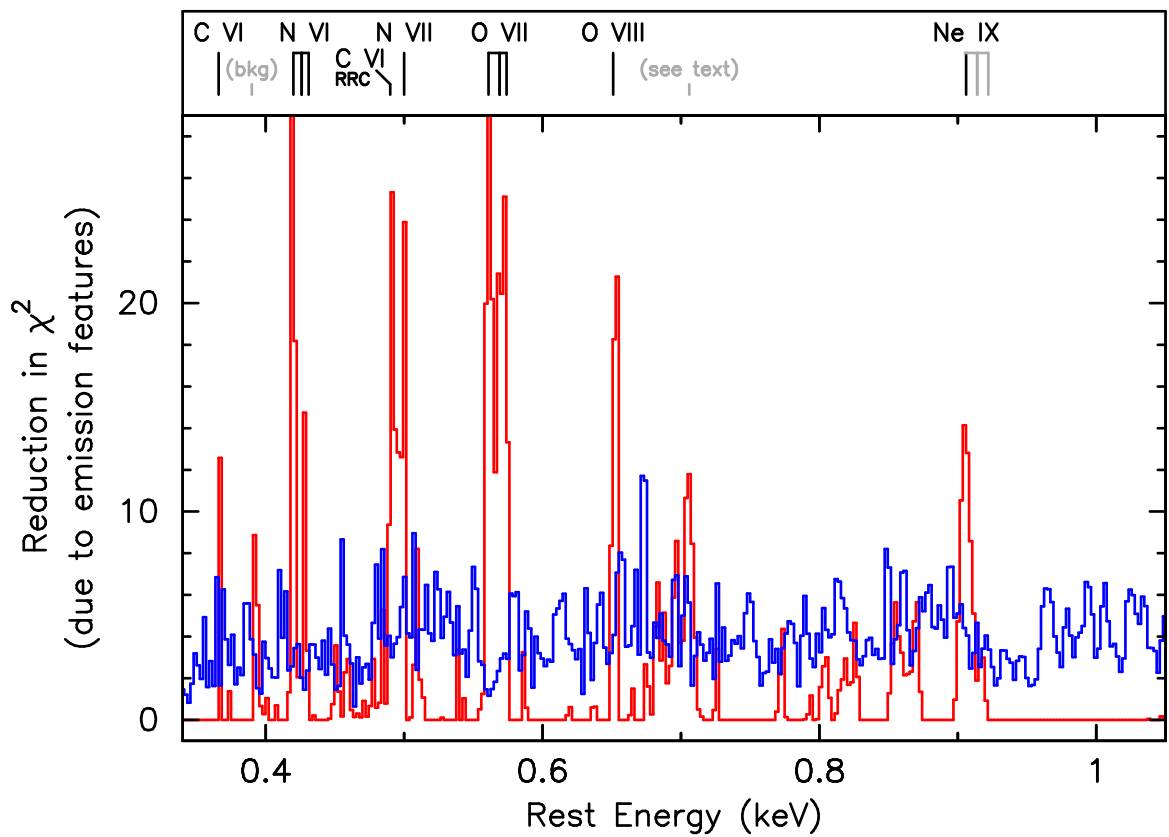}
\caption{Illustration of the significance of emission features. 
The red histogram shows 
the reduction in $\chi^2$ of the fit to the observed data 
when a narrow emission line is added to our model of the continuum 
at each energy. The data for the N{\sc vi} and O{\sc vii} lines are 
truncated; the peaks are actually at 31 and 87, respectively.
No significant emission features were found above 1~keV.
The blue histogram shows the maximum reduction in $\chi^2$ 
obtained at each energy when an identical 
analysis is performed on 20 sets of simulated spectra (see text).
This histogram therefore represents the 95\% confidence threshold
for narrow emission features.
The lines listed in Table~1 are marked, along with other 
features discussed in the text. It should be stressed that 
this plot conveys the change in statistic, 
{\it not} the intensity of the putative line.}
\end{figure}

\end{document}